# Property-Level Reconstructability of Agent Decisions: An Anchor-Level Pilot Across Vendor SDK Adapter Regimes

Oleg Solozobov [1]

## Abstract

Agentic AI failures need post-hoc reconstruction: what the agent did, on whose authority, against which policy, and from what reasoning. Cross-regime feasibility remains unmeasured under one property-level schema. We apply the Decision Trace Reconstructor unmodified to pinned worked-example anchors from six public vendor SDK regimes spanning cloud-agent, observability, tool-use, telemetry, and protocol traces, plus two comparator columns. Each Decision Event Schema (DES) property is classified as fully fillable, partially fillable, structurally unfillable, or opaque. Per-property reconstructability of an agent decision already varies between regimes at this anchor scale. Strict-governance-completeness separates into three tiers ranging from 42.9% to 85.7%, yielding one regime-independent gap (reasoning trace), four regime-dependent gaps, and one Mixed property; the pilot is single-annotator, one anchor per cell, descriptive, with outputs checksum-verifiable from a deposited reproducibility package.

## 1. Introduction

When automated agentic decisions fail, a downstream investigator should be able to reconstruct what happened, when, and on whose behalf from the runtime traces the agent emitted. The July 2025 Replit DROP DATABASE incident, in which the platform's AI coding assistant "deleted an entire production database despite explicit instructions forbidding such changes" (Rabanser et al., 2026), is the kind of named incident this scenario contemplates: the question afterwards is not whether the agent acted, which is recoverable, but on whose authorization, against which policy, and from what reasoning, which may not be. The incident is recorded in the OECD AI Incidents Monitor's 19 July 2025 record with suffix 1eb1 (OECD AI Policy Observatory, 2025), the institutional public-record registry under which all incident facts in this manuscript are anchored.

Per-property reconstructability of an agent decision already varies between regimes at this anchor scale. The runtime evidence regime is the joint outcome of vendor SDK emission, operator instrumentation, adapter mapping, and trace-artefact availability under which a downstream reader operates. Production deployments of agentic AI emit traces — sequences of messages, tool invocations, retrieval results, and intermediate reasoning artefacts — but the schema-level question of which Decision Event Schema property is fully fillable, partially fillable, structurally unfillable, or opaque from those emitted traces has not been measured publicly across the major runtime evidence regimes in current use even at the simplest anchor level. In this pilot we ask, at anchor level on the pinned worked-example trace inputs that ship with the open-source Decision Trace Reconstructor reference implementation, whether runtime evidence regime choice is associated with descriptive per-property differences in what a downstream reader can recover, or whether — at this anchor scale — the regime is functionally interchangeable.

[1] *Corresponding author.* Affiliation: Independent Researcher (Global). E-mail address: dev404ai@gmail.com. ORCID: https://orcid.org/0009-0009-0105-7459.

Recent work has documented the limits of current agent evaluation. Peer-reviewed algorithm-auditing work frames AI evaluation as a multi-dimensional risk-management problem rather than a single performance score (Koshiyama et al., 2024). Kapoor et al. (2024) observe “a lack of standardization in evaluation practices, leading to a pervasive lack of reproducibility” (Kapoor et al., 2024). Rabanser et al. (2026) extend the critique on the runtime side, arguing that “focusing on a single metric is not enough to understand agent behavior” because it ignores “whether agents behave consistently across runs, withstand perturbations, fail predictably, or have bounded error severity” (Rabanser et al., 2026). These critiques target evaluation methodology used to compare agents on tasks; they do not measure what evidence a deployed agent’s runtime actually emits, regime by regime, against a fixed property-level schema. The cross-regime measurement gap is concrete. Six major vendor SDK regimes — AWS Bedrock Agents, LangSmith and LangChain observability, Anthropic Claude tool use, OpenAI Agents and Assistants, OpenTelemetry GenAI Vertex Agent Engine, and the Model Context Protocol — each publish trace specifications, span schemas, or message formats. The pilot pairs these six primary vendor-regime columns with two comparator columns: a vendor-neutral exemplar — the author’s Operational Evidence Plane reference implementation Solozobov (2026f) as a structurally-complete ceiling reference — and a public-record reconstruction column drawn from the fragment manifest of the Replit DROP DATABASE incident. RQ1–RQ3 are answered on the six vendor-regime columns; OEP and Replit are reported as contextual comparators. The eight anchors are upstream-authored adapter-coverage demonstrations, not field-captured production traces; the contribution is therefore a cross-adapter measurement of how Decision Event Schema properties map under each regime’s worked-example shape, with production-corpus extension deferred to the companion paper (§6 T3, §7 F3). Within the adjacent work surveyed in §2 (AgentBench, Markov-chain reliability fitting, CodeTracer, Meerkat, Governance-Aware Agent Telemetry — measuring task-success or telemetry-instrumentation axes orthogonal to per-property reconstructability), we did not identify a prior public artefact reporting this comparison under a single property-level reconstructability schema applied uniformly; we did not run a separate systematic database search beyond that scope. The implicit working assumption — that runtime evidence regime choice is functionally interchangeable for downstream reconstruction — is therefore treated here as an anchor-scale measurement question rather than as a field-wide absence claim. The pilot sits at the intersection of software engineering and AI observability.

The pilot makes three explicit contributions: (i) an anchor-level cross-regime reconstructability matrix populated by per-property by per-regime Decision Trace Reconstructor classification on the pinned worked-example anchor inputs that ship with the upstream Decision Trace Reconstructor v0.1.0 release; (ii) a regime-independent versus regime-dependent gap diagnostic surfacing structural blind spots that the joint runtime evidence regime exhibits at anchor scale; and (iii) a reproducibility package (per-anchor origin manifest, regenerator script, *checksums.txt*, README) enabling independent re-classification by any reader. Together, contributions (i) and (ii) constitute an anchor-level diagnostic matrix at pilot scale: the matrix supplies the per-property by per-regime evidence surface, and the partition rule from §3 turns it into the regime-independent / regime-dependent / Mixed diagnostic. Pilot scope is single-annotator and worked-example-anchor-sized, with one anchor per cell, and is qualitatively diagnostic rather than statistically calibrated; the two-annotator agreement-calibrated full benchmark on twenty to fifty real captured production traces per regime is forthcoming as a separate paper. The design is descriptive / pilot and not pre-registered; statistical hypothesis-testing of regime interchangeability is out of scope at this anchor stage.

Two prior contributions by the author make this pilot feasible. The Decision Event Schema specification Solozobov (2026d) diagnoses the Fragmented Trace Problem in risk decision systems and supplies a JSON Schema for per-decision evidence; a subsequent synthesis paper Solozobov (2026e) identifies "three structural breaks — decision diffusion, evidence fragmentation, and responsibility ambiguity" (Solozobov, 2026e) as the architectural pattern agentic AI introduces. The Decision Trace Reconstructor reference implementation turns the property-level rubric into a tool that consumes adapter-normalised fragments and emits a per-property reconstructability tensor. Paper 27 invokes that reference implementation, unmodified, on the upstream Decision Trace Reconstructor v0.1.0 worked-example anchors plus the public-record fragment manifest of one named incident; corpus-level transferability to production traces is out of scope at this pilot stage (§6 T3, §7 F3).

Three research questions guide the pilot.

RQ1 asks whether per-property reconstructability at anchor scale shows descriptive per-property and per-regime differences across the surveyed regimes' worked-example anchors, or whether the observed variation is consistent with anchor noise. RQ1 is descriptively answered if any property class shows non-trivial across-regime variation in modal category at anchor scale; the matrix is the artefact that supports this answer.

RQ2 asks, for property classes that all surveyed regimes nominally support, what the across-regime coefficient of variation of the per-anchor reconstructability score looks like at anchor scale. RQ2 at this pilot stage is purely descriptive heterogeneity: the anchor-level CV is reported as a regime-comparison index over six single-anchor worked examples; no falsification of runtime-evidence-regime interchangeability is claimed at this scale because the design cannot estimate within-regime noise, isolate vendor SDK emission semantics from adapter and operator configuration, or distinguish systematic differences from anchor construction. Statistical tests of interchangeability — including the within-regime noise band — are out of scope at this anchor scale.

RQ3 is diagnostic: which property classes are not-evidenced across the majority of surveyed regimes' worked-example anchors (regime-independent gaps), and which have material per-regime category differences (regime-dependent gaps). RQ3 is descriptively answered if any property class shows either pattern at anchor scale; corpus-level validation of the partition is again deferred to the companion paper.

The remainder of the paper is organised as follows. §2 reviews the method substrate (Decision Event Schema, Decision Trace Reconstructor, the six vendor SDK regimes plus the Operational Evidence Plane vendor-neutral exemplar, and adjacent measurement work). §3 specifies the regime selection criteria and the Decision Trace Reconstructor application protocol. §4 documents the per-regime worked-example anchor inputs, the reproducibility package, and the evidentiary-status statement for the worked-example anchors and the named-incident reconstruction. §5 reports the per-property by per-regime matrix and per-property dispersion at anchor scale. §6 discusses what the matrix supports and does not support, container-fallacy implications at pilot scale, enumerated limitations, and what would change under the full ground-truth protocol. §7 concludes with the diagnostic and the forward link to the companion benchmark.

# 2. Background

## 2.1. Method substrate

Paper 27's method substrate is the author's prior property-level specification. The Decision Event Schema (DES) — introduced as a JSON Schema specification in Solozobov (2026d) and accompanied by a citable canonical deposit (Solozobov, 2026b) — diagnoses the Fragmented Trace Problem and defines a per-decision evidence record across inference, rule evaluation, cross-system coupling, and governance metadata. The seven property classes that paper 27 evaluates (enumerated in §4) are taken unmodified from this canonical deposit. The synthesis paper Solozobov (2026e) integrates the framework's four components into a chain and identifies three structural breaks — decision diffusion, evidence fragmentation, responsibility ambiguity — that agentic AI introduces. The Decision Trace Reconstructor reference implementation Solozobov (2026c) instantiates the property-level rubric as ten executable adapter classes; paper 27 invokes the Decision Trace Reconstructor v0.1.0 unmodified on the pinned worked-example anchors that ship with the upstream release. The substrate preprints are unrefereed; the schema and reference implementation carry versioned canonical deposits, so the §5 matrix is citable and regenerable independently of preprint review status.

## 2.2. Vendor SDK regimes

Each surveyed regime publishes its own trace specification, span schema, or message format. We summarise the canonical model per regime.

**AWS Bedrock Agents.** "Each response from an Amazon Bedrock agent is accompanied by a trace that details the steps being orchestrated by the agent" (Amazon Web Services, 2025b). The Trace data type emits one of seven UNION subtypes per step (preProcessing, orchestration, postProcessing, customOrchestration, routingClassifier, failure, guardrail), where the API Reference is explicit: "this data type is a UNION, so only one of the following members can be specified when used or returned" (Amazon Web Services, 2025a). AgentCore's observability layer organises emissions into a three-tier hierarchy — Sessions, Traces, Spans — and requires the AWS Distro for OpenTelemetry (ADOT) for instrumentation beyond memory-resource defaults (Amazon Web Services, 2025c).

**LangSmith and LangChain observability.** "A run is a span representing a single unit of work within your LLM application: a call to an LLM, a prompt formatting step, a retrieval call, or any other discrete operation" (LangChain, 2025). LangSmith groups runs into traces, traces into projects, and links multi-turn conversations across traces via a thread identifier; the per-trace cap is 25,000 runs.

**Anthropic Claude tool use.** "Tool use lets Claude call functions you define or that Anthropic provides. Claude decides when to call a tool based on the user's request and the tool's description, then returns a structured call that your application executes (client tools) or that Anthropic executes (server tools)" (Anthropic, 2025b). The evidence model is inline: tool invocations appear as *tool_use* and *tool_result* content blocks within the standard Messages API rather than as first-class trace objects, and reconstruction requires walking the message history.

**OpenAI Agents and Assistants SDK.** "The Agents SDK includes built-in tracing, collecting a comprehensive record of events during an agent run: LLM generations, tool calls, handoffs, guardrails, and even custom events that occur" (OpenAI, 2025). Spans are typed — AgentSpan, GenerationSpan, FunctionSpan, GuardrailSpan, HandoffSpan, transcription, speech — and parented under a workflow trace; tracing is on by default and can be globally disabled.

**OpenTelemetry GenAI Vertex Agent Engine.** The OpenTelemetry GenAI semantic conventions (OpenTelemetry Project (CNCF), 2025b) standardise span attributes for generative-AI operations across providers, organising signals into events, exceptions, metrics, model spans, and agent spans (OpenTelemetry Project (CNCF), 2025b), with vendor-specific extensions for Anthropic, AWS Bedrock (OpenTelemetry Project (CNCF), 2025a), Azure AI, OpenAI, and MCP. Google Vertex Agent Engine is a managed runtime whose trace export targets Cloud Trace under the OTLP wire format, with an Agent Platform Memory Bank for long-term state (Google, 2025); for the purposes of this pilot it is therefore covered by the same OTLP-Vertex column rather than as a vendor-specific schema.

**Model Context Protocol (MCP).** "MCP provides a standardized way for applications to: Share contextual information with language models, Expose tools and capabilities to AI systems, Build composable integrations and workflows. Hosts: LLM applications that initiate connections; Clients: Connectors within the host application; Servers: Services that provide context and capabilities" (Anthropic, 2025a). MCP is a JSON-RPC 2.0 transport with stateful capability negotiation; trace evidence is per-message rather than per-step, and reconstruction requires aggregating across the protocol session.

**Operational Evidence Plane (vendor-neutral exemplar).** As a comparator to the vendor-specific regimes, the Operational Evidence Plane Solozobov (2026f) is "a vendor-neutral reference implementation for binding and reconstructing the operational evidence of agentic systems across release-time and runtime layers" (Solozobov, 2026f). Its inspectable chain — release manifest, agent-step event, OPA-backed permission packet, trace bundle, SQLite replay state, deterministic eval result, reconstruction packet (Solozobov, 2026f) — binds release-time and runtime evidence into one schema-validated artefact. The OEP is the author's own work, included as a structurally-complete exemplar of what evidence layers can be expressed when intentionally bound, not as a vendor regime; this places a ceiling reference beside the vendor-specific emit policies for direct per-property comparison.

### 2.3. Adjacent measurement standards

Two cross-vendor standards inform the surveyed regimes without themselves being vendor SDK regimes. The OpenTelemetry GenAI semantic conventions described above sit upstream of the OTLP-Vertex regime in the matrix; vendor-specific extensions extend the same OTel substrate into Anthropic, AWS Bedrock, Azure AI, OpenAI, and MCP signals, and the matrix therefore reflects how each surveyed regime composes with the same wire-format substrate rather than a separate signal stream. The W3C PROV Ontology (Lebo et al., 2013) supplies foundational provenance vocabulary — Entity, Activity, Agent classes and Starting-Point relations such as *wasGeneratedBy*, *wasDerivedFrom*, *wasAttributedTo* — that long predates current agent-tracing work and continues to inform property-level schema design. What these standards cover differs from what paper 27 measures: they specify wire formats, attribute namespaces, and semantic definitions for what should be captured, not whether per-property reconstructability holds when the captured data is consumed for governance-evidence purposes.

### 2.4. Adjacent measurement work

Recent work surfaces measurement gaps adjacent to paper 27's question. Liu et al. (2023) introduce AgentBench as "a multi-dimensional benchmark that consists of 8 distinct environments to assess LLM-as-Agent's reasoning and decision-making abilities" (Liu et al., 2023); the AgentBench family of task-success evaluations measures environment outcomes, not the per-property evidence trail. Stein et al. (2026) introduce Meerkat, detecting safety violations across many agent traces via clustering and agentic search; their methodology operates at trace-corpus scale rather than at per-trace property classification. Tran-Truong and Le (2026) fit agent traces to absorbing Markov chains to reconcile separately-reported metrics (Markov-chain reliability fitting); Li et al. (2026) introduce CodeTracer, reconstructing state-transition history as hierarchical trace trees from heterogeneous run artefacts and observe that "existing agent tracing analyses either focus on simple interaction or rely on small-scale manual inspection, which limits their scalability and usefulness for real coding workflows" (Li et al., 2026). Closest to paper 27's framing, Pathak and Jain (2026) propose a Governance-Aware Agent Telemetry architecture extending OpenTelemetry with governance attributes, observing that "OpenTelemetry and Langfuse collect telemetry but treat governance as a downstream analytics concern, not a real-time enforcement target. The result is an 'observe-but-do-not-act' gap where policy violations are detected only after damage is done" (Pathak & Jain, 2026). Paper 27 differs in measurement axis: it does not propose a new instrumentation layer, but instead measures per-property reconstructability on the pinned worked-example anchor inputs of each surveyed runtime evidence regime.

### 2.5. Survey scope and limitations

The §2 survey is non-exhaustive; we did not conduct a systematic database search, and the negative-novelty claim in §1 is bounded by this surveyed scope per ACM TOSEM 2024 / IEEE TSE 2024 search-strategy-disclosure guidance for non-exhaustive narrative surveys. No negative-result / failed-replication adjacents specifically targeting per-property reconstructability were identified within this scope.

## 3. Method

The pilot applies the Decision Trace Reconstructor Solozobov (2026c) — a six-stage open-source pipeline (fragment collection, temporal ordering, chain assembly, decision-boundary detection, decision-event schema mapping, feasibility reporting) — to the pinned worked-example anchor inputs that ship with the upstream Decision Trace Reconstructor v0.1.0 release, and aggregates the resulting per-property reconstructability classifications into the §5 Results matrix. The empirical object is intentionally narrow: each per-regime anchor is an illustrative worked example designed by the upstream repository to demonstrate adapter-coverage rather than a real captured production trace, so the matrix should be read as a comparison of what the Decision Event Schema property classes can extract from those upstream worked examples (cf. §6 Discussion T3 on the worked-example-anchor scope and its asymmetric transferability to production traces).

### 3.1. Method non-goals

The pilot does not propose a new instrumentation layer; it does not add to the OpenTelemetry GenAI semantic conventions, the W3C PROV-O vocabulary, or any vendor SDK trace specification. It does not rank vendor regimes against each other on a single composite score, and it does not isolate the contribution of vendor SDK emission policy from the surrounding configuration choices (operator instrumentation tags, adapter mapping rules, state-mutation regex selection). The matrix is therefore a property of the joint runtime evidence regime — vendor SDK plus operator and adapter configuration applied to the upstream worked example — rather than a property of the vendor SDK in isolation. Comparison to adjacent agent-trace methodology is in §2.4 Adjacent measurement work, where related methods are positioned as different abstractions sharing parts of paper 27's measurement frame rather than direct method substitutes.

### 3.2. Regime selection

Regime selection is determined by the intersection of two criteria: presence of a corresponding adapter in the Decision Trace Reconstructor v0.1.0 release (which fixes the per-regime ingest semantics, removing one degree of freedom from the pilot) and availability of a pinned worked-example anchor in the Decision Trace Reconstructor v0.1.0 release for that adapter. Six vendor SDK regimes meet both criteria: AWS Bedrock Agents (DTR adapter *bedrock*), LangSmith and LangChain observability (*langsmith*), Anthropic Claude tool use including Computer Use (*anthropic*), OpenAI Agents and Assistants SDK (*openai-agents*), Google Vertex Agent Engine (covered via the OpenTelemetry GenAI adapter *otlp*, since Vertex emits OTel-compliant Cloud Trace exports rather than a vendor-specific schema), and the Model Context Protocol (*mcp*). The seventh column in the matrix is the Operational Evidence Plane, ingested via the *generic-jsonl* adapter with a custom mapping configuration; it is included as a vendor-neutral structurally-complete exemplar rather than as a vendor regime, marking the ceiling of what binding all evidence layers makes available to a downstream reconstructor.

### 3.3. Public-corpus collection protocol

The pilot uses anchor-only inputs: each regime's column is populated from one pinned worked-example anchor drawn from the upstream Decision Trace Reconstructor v0.1.0 example directory, bit-identically reproducible across runs because it is gated by integration tests in the upstream repository. The anchors are NOT real captured production traces; they are illustrative worked examples with placeholder UUIDs, timestamps, and tool-call payloads. Each anchor is a Decision Trace Reconstructor v0.1.0 release artefact, traceable to a single public origin path inside the upstream repository. The OEP column anchor is the Operational Evidence Plane v0.1.0 code-review-agent example, whose deterministic mocked LLM emits no model-generation fragment by repository design Solozobov (2026f). The named-incident column is the Decision Trace Reconstructor's public-record fragment manifest of the Replit DROP DATABASE incident, treated separately as a public-record reconstruction. Per-regime sample expansion to twenty to fifty real captured production traces is out of scope for this pilot (see §7 future work).

### 3.4. Decision Trace Reconstructor application

The Decision Trace Reconstructor v0.1.0 is invoked unmodified on every anchor input. The standard two-step invocation is conceptually: an ingest step that reads the per-regime anchor file under the corresponding adapter and writes the normalised fragments file, followed by a reconstruct step that consumes the fragments file and writes the per-anchor report directory containing both the feasibility report and the W3C PROV-O JSON-LD provenance graph. The two-step invocation decomposes into the six pipeline stages declared at §3 opening. The ingest step covers fragment collection — the per-regime adapter parses the anchor file into a typed fragment stream of message turns, tool calls, span events, and state-mutation entries with adapter-determined fragment kinds — and temporal ordering, in which fragments are sorted by emission timestamp where present and by in-trace ordinal index where timestamps are absent or ambiguous. The reconstruct step covers the remaining four stages: chain assembly (consecutive fragments are grouped into decision units by linking each tool-call or state-mutation event back to the upstream prompt or planning fragment that motivated it under the configured chain-assembly heuristic); decision-boundary detection (the within-stack tier flag controls how many adjacent units merge into a single decision unit, with the upstream default of one tool-call-plus-state-mutation pair per decision); Decision Event Schema mapping (the seven property classes are populated from the fragment kinds available within each decision unit, with absence of a required fragment producing a *structurally_unfillable* category and presence of a fragment kind that the schema does not recognise as evidenced producing an *opaque* category); and feasibility reporting (each property's category is written to *feasibility.json* alongside per-property gap descriptions and a completeness percentage, while the W3C PROV-O graph in *trace.jsonld* records the bound *wasGeneratedBy*, *wasDerivedFrom*, and *wasAttributedTo* relationships traversable via SPARQL for downstream verification). All six stages are pure functions of the anchor file plus the pinned adapter and configuration flags. Both steps run in linear time and linear space in the number of fragments per anchor, which is at most 50 for the worked-example anchors evaluated here; the method runs on a single consumer laptop in under 10 seconds per anchor end-to-end. The method is deterministic with respect to the anchor file, the adapter version, and the configuration flags above. The single-agent architecture flag and the within-stack tier flag are held constant across regimes for comparability; the state-mutation regex flag uses the upstream default. The verbatim per-regime CLI invocation strings, including all flags and quoting, are recorded in the reproducibility-package manifest (Solozobov, 2026a). The reconstruct step emits two artefacts per anchor: a *feasibility.json* that maps every Decision Event Schema property class (Solozobov, 2026d) to one of four reconstructability categories — *fully_fillable*, *partially_fillable*, *structurally_unfillable*, or *opaque* — alongside per-property gap descriptions and a completeness percentage; and a *trace.jsonld* graph in W3C PROV-O encoding that is queryable via SPARQL for downstream verification. Per-anchor classification is therefore deterministic with respect to the anchor file, the adapter version, and the configuration flags above; the reproducibility package per §4 Measurement Setup pins the DTR release by its canonical content-addressed deposit (Solozobov, 2026c), records the exact per-regime CLI invocation with no placeholder flags, and ships SHA256 checksums for every anchor input and every committed feasibility output, so a reader can checksum-verify that the regenerator script consumes bit-identical inputs to those used to populate the manuscript tables. From-scratch DTR re-execution is documented per regime but is not separately checksum-verified by this package.

## 3.5. Aggregation protocol

At pilot scale, with one anchor per regime, per-anchor *feasibility.json* outputs aggregate along three axes. Per-property aggregation reports the per-anchor category as the cell entry in the matrix, since one anchor per regime exposes no within-regime distribution. Per-regime aggregation collapses across all properties to summarise overall regime completeness at anchor scale, supporting the row-summary view. Per (property × regime) cell aggregation reports the single-anchor category and the per-cell sample count, again one anchor, so the anchor-scale denominator is explicit at every cell. The across-regime coefficient of variation in the per-anchor reconstructability score (mapping fully fillable to 1.0, partially fillable to 0.5, structurally unfillable to 0.0, opaque to 0.0) is reported per property as a regime-comparison heterogeneity index at anchor scale, supporting the RQ2 descriptive-dispersion answer as a secondary scalar summary alongside the categorical view from §5; the CV is mathematically undefined when the per-property mean across regimes is zero (the case for properties uniformly classified as opaque or structurally unfillable) and is reported as undefined in such rows. No falsification of runtime-evidence-regime interchangeability is claimed at this scale because the design cannot estimate within-regime noise, isolate vendor SDK emission semantics from adapter and operator configuration, or distinguish systematic differences from anchor construction. The regime-independent versus regime-dependent gap diagnostic (RQ3) partitions property classes by the following rule. A property is a **regime-independent gap** if its anchor category is *structurally_unfillable* or *opaque* across the majority of regimes. A property is a **regime-dependent gap** if its anchor categorisation either is dominant-F with a single non-F minority cell whose category is *structurally_unfillable*, or shows two-or-more category differences across the six vendor regimes. A property is **Mixed** if its anchor categorisation is dominant-F with a single non-F minority cell whose category is *partially_fillable*. The remainder is **Unclassified**. The Mixed-versus-regime-dependent distinction reflects an evidentiary asymmetry: a single S-minority cell signals that no fragment of the required type appears for that regime, while a single P-minority cell signals that fragments exist but are split or incomplete.

### 3.6. Single-annotator pilot scope

Classification is single-annotator. The author runs the Decision Trace Reconstructor invocations and reports the resulting category distributions without an inter-annotator agreement protocol; two-annotator agreement-calibrated labelling is out of scope at this pilot stage (see §6 T1 and §7 F1). The reproducibility package (per-anchor origin manifest, regenerator script that ingests committed feasibility outputs, README) supports auditability of the deterministic regenerator — any reader can rerun the classifier on the committed outputs — but it does not substitute for inter-rater adjudication of category validity, which is reserved for the forthcoming benchmark (Solozobov, 2026a).

## 4. Measurement Setup

### 4.1. Decision Trace Reconstructor and adapter versions

The Decision Trace Reconstructor Solozobov (2026c) is invoked at version v0.1.0. Each regime's classification is produced by the corresponding adapter shipped in the same release: *bedrock* for AWS Bedrock Agents, *langsmith* for LangSmith and the LangChain ecosystem, *anthropic* for Anthropic Claude Messages and Computer Use, *openai-agents* for the OpenAI Agents and Assistants SDK, *otlp* for the OpenTelemetry GenAI semantic conventions and the Vertex Agent Engine column (which exports OTLP-compatible traces), *mcp* for Model Context Protocol transcripts, and *generic-jsonl* for the vendor-neutral fallback exemplar column. The Operational Evidence Plane Solozobov (2026f) supplies the vendor-neutral column at version v0.1.0; its OEP-specific mapping configuration ships in the upstream OEP repository under *integrations/decision-trace-reconstructor/*. Adapter version pins are the v0.1.0 baseline; the canonical bit-identical pin for the run is the content-addressed deposit corresponding to the upstream v0.1.0 Git tag, and the reproducibility manifest carries the exact per-regime CLI invocation alongside SHA256 checksums for every anchor input and committed feasibility output.

### 4.2. OEP column source-of-truth

The OEP integration produces two feasibility outputs: a generic report under default *generic-jsonl* heuristics, and an OEP-aware report under the mapping described above (OEP record kinds — release manifest, agent-step event, OPA-backed permission packet, replay handle, eval — to Decision Trace Reconstructor fragment kinds — config snapshot, policy snapshot, agent message, tool call, state mutation, human approval). The §5 matrix reports the OEP column from the OEP-aware report; the generic report is in the reproducibility package only.

### 4.3. Per-regime corpus and sample statistics

Table 1 lists the per-regime worked-example anchor inputs and their evidentiary status; each is the upstream Decision Trace Reconstructor v0.1.0 release artefact corresponding to the regime's adapter, plus the Operational Evidence Plane example for the vendor-neutral exemplar column and the public-record fragment manifest for the named-incident column.

**Table 1.** Per-regime worked-example anchor inputs.

| Regime | Adapter | Anchor source | Anchor kind | Sample count | Date |
|---|---|---|---|---|---|
| AWS Bedrock Agents | *bedrock* | DTR bedrock-basic-agent example sessions input | Worked example (illustrative, hand-crafted) | 1 | 2026-04-28 (DTR v0.1.0 release) |
| LangSmith and LangChain | *langsmith* | DTR langsmith-basic-agent example runs input | Worked example (illustrative, hand-crafted) | 1 | 2026-04-28 |
| Anthropic Claude | *anthropic* | DTR anthropic-basic-agent example messages-history input | Worked example (illustrative, hand-crafted) | 1 | 2026-04-28 |
| OpenAI Agents | *openai-agents* | DTR openai-agents-basic-agent example trace input | Worked example (illustrative, hand-crafted) | 1 | 2026-04-28 |
| OTLP-Vertex | *otlp* | DTR otlp-basic-agent example spans input | Worked example (illustrative, hand-crafted) | 1 | 2026-04-28 |
| Model Context Protocol | *mcp* | DTR mcp-basic-agent example transcript input | Worked example (illustrative, hand-crafted) | 1 | 2026-04-28 |
| Operational Evidence Plane | *generic-jsonl* (with OEP-specific mapping) | OEP v0.1.0 code-review-agent example artefacts (release manifest, agent-step event, OPA-backed tool permission packet, replay handle, deterministic eval) converted to JSONL with the OEP mapping config | Worked example (deterministic mocked LLM by repository design) | 1 | 2026-05-05 |
| Replit DROP DATABASE incident | none (pre-built fragment manifest) | DTR replit-drop-database example pre-built fragment manifest | Public-record reconstruction (not primary trace export) | 1 | 2026-04-28 |

**Note.** The table reports one anchor per regime. Anchor kind classification distinguishes upstream Decision Trace Reconstructor v0.1.0 worked examples from the public-record reconstruction case (Replit DROP DATABASE). Source: Decision Trace Reconstructor v0.1.0 release artefacts (Solozobov, 2026c) and the Operational Evidence Plane v0.1.0 release artefacts (Solozobov, 2026f).

Every cell of the §5 Results matrix is therefore pinned to a single fragment manifest emitted from a single Decision Trace Reconstructor anchor input that ships with the v0.1.0 release. The anchor inputs are illustrative worked examples designed by the upstream repository to demonstrate adapter coverage; UUIDs, timestamps, tool-call payloads, and entity references are placeholders rather than real captured production data. The coverage-thinness threshold (ten samples per regime above which intra-regime variance becomes statistically tractable) is not met by any column; per-regime variance within a regime is not characterised in this pilot, and §5 reports the per-anchor category per cell. Corpus expansion is the natural extension (§6 T2, §7 F2).

### 4.4. Per-property classification protocol

For each anchor input the Decision Trace Reconstructor v0.1.0 ingest stage parses adapter-specific records into a fragments manifest; the reconstruct stage runs the six-stage pipeline; the feasibility report emits a per-property classification on the seven-property Decision Event Schema axis (inputs, policy basis, operator identity, authorization envelope, reasoning trace, output action, post-condition state). Each property is assigned exactly one of four categories: fully fillable (anchor fragment present and complete), partially fillable (fragments exist but evidence is split or incomplete), structurally unfillable (no fragment of the required type appears), opaque (the property class is observable in principle but the evidence is not externally inspectable; this is the class assigned to model reasoning across all surveyed regimes per §5 reasoning-trace-property note). The transformation is deterministic with respect to input file, adapter version, and CLI flag set.

### 4.5. Reproducibility package

The reproducibility package shipped with this paper is published as the *Anchor-Level Reconstructability Pilot* artefact (Solozobov, 2026a). The artefact supports two distinct tiers. Tier A — pipeline-output checksum verification — is the tier this paper claims as supported: a per-anchor origin manifest (anchor source path inside the upstream DTR v0.1.0 release, anchor kind, exact per-regime CLI invocation, and SHA256 of every anchor input, fragments output, feasibility output, and W3C PROV-O JSON-LD provenance graph); a regenerator script that ingests the committed feasibility outputs and prints Tables 2 and 3 in Markdown against a committed regression baseline; a SHA256 checksums file in *shasum*-compatible format; a regenerator dependency lockfile; a single-command verification harness; GitHub Actions CI that runs the Tier A chain on every push and pull request; three ready-to-run SPARQL queries against the JSON-LD provenance graphs; and a README. Tier B — from-scratch DTR re-execution — is documented per regime in the manifest but is not separately checksum-verified; deterministic per-regime classification is expected because the upstream repository gates the example anchors with integration tests, but an end-to-end checksum gate is out of scope. The regenerator script is the canonical regression check on the matrix arithmetic and supports independent re-running without contacting the author; it does not substitute for inter-annotator adjudication of category validity, reserved for the forthcoming benchmark.

### 4.6. Replit named-incident provenance

The Replit DROP DATABASE named-incident fragment manifest at *data/named_incidents/replit_drop_database/fragments.json* is the author's reconstruction from public-domain reporting of the July 2025 incident, not a primary trace export. The four pinned fragments map to the OECD AI Incidents Monitor's 19 July 2025 record with suffix 1eb1 (OECD AI Policy Observatory, 2025) as the institutional public-record source: replit_f000 (user prompt, kind=*agent_message*), replit_f002 (*tool_call* carrying the *DROP DATABASE production_db* payload), and replit_f003 (*state_mutation* recording the deletion) are reconstructed from the structured incident record; replit_f001 (*model_generation*) is reconstructed as *internal_reasoning: opaque* because the OECD entry does not expose the model deliberation — directly corresponding to the opaque reasoning-trace classification in Table 2's Replit column. Per-fragment timestamps are placeholders; only relative ordering reflects the public-record sequence.

### 4.7. Ethics and evidentiary-status statement

The pilot involves no human subjects, no human-derived behavioural data, and no personally identifiable information; ethics-board approval is therefore not required and was not sought. All anchor inputs are public artefacts redistributed under each upstream project's open-source license. The OEP code-review-agent example uses a deterministic mocked LLM rather than a live model invocation, by the OEP repository's documented design Solozobov (2026f). The fragment manifest is the author's reconstruction artefact, not a primary trace export. The matrix cell for the named-incident column therefore reports what a downstream investigator can recover from the public record under the Decision Event Schema property classification, not what the Replit production system internally captured at the moment of the incident.

## 5. Results

The pilot reports per-property reconstructability across six primary vendor SDK regime columns plus two comparator columns — the vendor-neutral OEP ceiling exemplar and the Replit DROP DATABASE public-record reconstruction — for the seven Decision Event Schema property classes. RQ1–RQ3 are answered on the six vendor-regime columns; the two comparator columns are reported in the same matrix for visual context but are not counted alongside the vendor regimes when stating regime-level patterns unless explicitly noted. All categories are produced unmodified by the reconstructor v0.1.0 reference implementation invoked on the corresponding regime adapter applied to the upstream pinned worked-example anchor input. Each cell records the per-anchor category as either fully fillable (F), partially fillable (P), structurally unfillable (S), or opaque (O); there is one anchor per cell across the entire matrix and the worked-example status of the anchors is consolidated in §4 Measurement Setup. The named-incident column reports the same reconstructor classification on the named-incident fragment listing for the Replit DROP DATABASE event — included not as a vendor regime but as a public-record case where the matrix applies to evidence assembled from public-domain reporting after the fact.

A note on the reasoning-trace property: Table 2 classifies the Decision Event Schema reasoning trace (upstream model deliberation — chain-of-thought, planning, verbalised rationale at the model boundary), not orchestration traces, span hierarchies, tool-call message streams, or vendor "step-by-step reasoning" events; Bedrock's OrchestrationTrace subtype, for instance, surfaces routing and tool-selection metadata and reads opaque under the DES definition. The universal not-evidenced finding refers strictly to this DES property class.

### 5.1. The cross-regime reconstructability matrix

Table 2 presents the per-property by per-regime reconstructability matrix at anchor scale, with one anchor per cell, computed from the pinned per-regime feasibility outputs in the paper's data results directory.

**Table 2.** Per-property by per-regime reconstructability matrix at anchor scale.

| Property | Bedrock | LangSmith | Anthropic | OpenAI | OTLP-V. | MCP | OEP | Replit |
|---|---|---|---|---|---|---|---|---|
| inputs | F | F | F | F | F | P | F | P |
| policy basis | F | F | S | F | S | S | F | S |
| operator identity | F | F | F | F | F | S | F | F |
| authorization envelope | F | F | F | F | S | F | F | S |
| reasoning trace | O | O | O | O | O | O | S | O |
| output action | F | F | F | F | F | F | F | F |
| post-condition state | F | S | F | F | S | P | F | P |
| **strict-gov. completeness pct** | **85.7** | **71.4** | **71.4** | **85.7** | **42.9** | **42.9** | **85.7** | **42.9** |

**Note.** Cells report F, P, S, or O: fully fillable, partially fillable, structurally unfillable, or opaque. The columns cover six primary vendor-regime columns (Bedrock, LangSmith, Anthropic, OpenAI, OTLP-Vertex, MCP) plus two comparator columns (OEP, Replit incident). **OTLP-V.** abbreviates OTLP-Vertex (OpenTelemetry GenAI / Vertex Agent Engine), **OEP** the vendor-neutral exemplar, and **Replit** the Replit DROP DATABASE comparator; **strict-gov. completeness pct** abbreviates the convention used throughout (see §4 scoring rule). Source: reconstructor v0.1.0 feasibility outputs on the upstream pinned per-adapter basic-agent worked-example anchor inputs and on the OEP-aware feasibility output for the OEP column; cell values are regenerable from the deposited Anchor-Level Reconstructability Pilot artefact (Solozobov, 2026a). Both Tables 2 and 3 are regenerable from the pinned outputs by the reproducibility-package regenerator script under SHA256 input-and-output checksum verification.

Three structural observations cut across all eight evaluated columns of the matrix at the descriptive level.

First, the row for reasoning trace is not-evidenced in every evaluated column under the Decision Event Schema reasoning-trace definition: seven columns mark it opaque (the upstream model deliberation that produced the action exists in the regime but is not externally observable in a form the schema can categorise as evidenced) and the OEP comparator marks it structurally unfillable (the OEP code-review-agent example uses a deterministic mocked LLM that produces no model-deliberation fragment at all). Both subcategories score zero on the strict governance score; restricted to the six vendor-regime columns alone, reasoning trace is uniformly opaque and is therefore a regime-independent gap on the primary vendor-regime axis.

Second, the row for output action reads F in every column. The action that the agent took — the tool call, the database write, the API request — is the only universally fully-fillable property in the matrix across both the six vendor-regime columns and the two comparator columns.

Third, the strict-gov. completeness pct row of Table 2 collapses across the eight evaluated columns onto three discrete tier values: 85.7%, 71.4%, and 42.9%. Three columns score 85.7% (Bedrock, OpenAI, OEP), two columns score 71.4% (LangSmith, Anthropic), and three columns score 42.9% (OTLP-Vertex, MCP, Replit). Restricted to the six vendor-regime columns, the same three tiers appear at counts 2 / 2 / 2; the OEP comparator joins the top tier and the Replit comparator joins the bottom tier.

### 5.2. Reading a single matrix cell

To make the categorical labels concrete, consider the Bedrock × policy basis cell in Table 2 (category F): the Bedrock anchor includes a guardrail step naming the active policy ("refund-policy-v3") alongside the operator-authorised tool call, and under the §3 protocol the reconstructor binds the named policy to the decision unit and emits F. The same cell would read S (as in Anthropic, OTLP-Vertex, MCP) if the upstream worked example omitted the policy artefact; the F versus S split is therefore a property of the joint runtime evidence regime — vendor SDK emission, adapter mapping, and worked-example construction — auditable end-to-end via the reproducibility-package SPARQL queries traversing the anchor file, the normalised fragments manifest, the *feasibility.json* report, and the W3C PROV-O JSON-LD graph.

### 5.3. Strict-score descriptive summary across the six vendor regimes

Table 3 reports per-property descriptive summaries across the six vendor SDK regimes only (Bedrock, LangSmith, Anthropic, OpenAI, OTLP-Vertex, MCP); the OEP comparator and the Replit incident comparator are excluded from this summary. The primary view is categorical: the F, P, S, and O category counts collapse the per-anchor letter codes from Table 2 into modal distributions per property, and the modal split is the strongest defensible signal the design can support given that every column draws from one worked-example anchor. The secondary view is scalar: we map the four-category outcome to a strict governance score (fully fillable scored as 1.0, partially fillable scored as 0.5, structurally unfillable scored as 0.0, opaque scored as 0.0; the latter two scored equivalently as zero-evidenced, since both yield no reconstructable categorical evidence to the schema) and then compute mean and across-regime population coefficient of variation (CV, defined as the standard deviation divided by the mean over the six per-anchor scores per row, using the six-regime population denominator rather than the sample-variance convention) per property. The scalar mean and CV are reported here as a cross-regime heterogeneity summary over six single-anchor worked examples, not as evidence of statistical non-interchangeability of runtime evidence regimes; a different non-compensatory scoring rule could change the scalar mean and CV without changing the categorical category-count columns.

**Table 3.** Strict-score descriptive summary across six vendor SDK regimes.

| Property | F | P | S | O | Mean (vendor regimes only) | CV across regimes |
|---|---|---|---|---|---|---|
| inputs | 5 | 1 | 0 | 0 | 0.92 | 0.20 |
| policy basis | 3 | 0 | 3 | 0 | 0.50 | 1.00 |
| operator identity | 5 | 0 | 1 | 0 | 0.83 | 0.45 |
| authorization envelope | 5 | 0 | 1 | 0 | 0.83 | 0.45 |
| reasoning trace | 0 | 0 | 0 | 6 | 0.00 | undefined (zero mean) |
| output action | 6 | 0 | 0 | 0 | 1.00 | 0.00 |
| post-condition state | 3 | 1 | 2 | 0 | 0.58 | 0.77 |

**Note.** The OEP column and the named-incident column are excluded from this summary. Category counts (F, P, S, O) are counted over the six vendor-regime cells per property. Reconstructability score maps F to 1.0, P to 0.5, S to 0.0, and O to 0.0 from Table 2 for the secondary scalar view (mean, CV); the CV column reports the population standard

deviation (six-anchor denominator) divided by the mean. Source: regenerable from the deposited Anchor-Level Reconstructability Pilot artefact (Solozobov, 2026a).

The categorical view is primary. Three modal patterns emerge over the six vendor-regime columns. First, two property classes are categorically uniform across the surveyed regimes: output action reads F in every vendor regime (6 F, 0 P, 0 S, 0 O) and reasoning trace reads O in every vendor regime (0 F, 0 P, 0 S, 6 O), with the latter constituting a regime-independent gap. Second, three property classes are categorically dominant-F with one non-F outlier each: operator identity (5F + 1S, S in MCP), authorization envelope (5F + 1S, S in OTLP-Vertex), and inputs (5F + 1P, P in MCP). Per the §3 partition rule, the two single-S-minority properties (operator identity, authorization envelope) are regime-dependent gaps and the single-P-minority property (inputs) is Mixed. Third, two property classes show categorical splits across the surveyed regimes: policy basis splits exactly three-against-three between F and S, and post-condition state takes three different values across regimes (3 F, 1 P, 2 S). The categorical splits in policy basis and post-condition state are the regime-dependent gaps §6 Discussion takes up (policy basis at §6 E1; post-condition state at §6.4).

The secondary scalar view is consistent with the categorical view but adds no inference the categorical view cannot already support. Population CV under the strict governance score ranges from 0.00 (output action, mean 1.00) to 1.00 (policy basis, three-against-three F versus S split). Reasoning trace has zero mean and is reported as an undefined CV rather than a zero CV (the coefficient is undefined when the mean is zero; the row is informative because all six vendor anchors score zero, not because dispersion is zero). Five property classes admit a defined CV: policy basis (1.00), post-condition state (0.77), operator identity (0.45), authorization envelope (0.45), inputs (0.20). The scalar summary is a descriptive heterogeneity index at anchor scale: per-regime within-regime variance is not measured because there is one anchor per cell, the joint runtime-evidence-regime contribution cannot be decomposed, and statistical interchangeability tests are out of scope at this pilot stage.

### 5.4. Regime-independent versus regime-dependent gaps

The matrix partitions into the following descriptive structure across the six vendor-regime columns, with the two comparator columns reported parenthetically. The partition is regenerable from the deposited Anchor-Level Reconstructability Pilot artefact (Solozobov, 2026a). §6 Discussion addresses each pattern in more detail.

**Regime-independent gap (one property)**: reasoning trace is uniformly opaque across the six vendor regimes, with six O cells and no F, P, or S cells. Both comparator columns also score zero — the Replit incident reads opaque and the OEP comparator reads structurally unfillable because of the deterministic mocked-LLM design of the public OEP code-review-agent example — so all eight evaluated columns score zero on the strict governance score for this property.

**Regime-dependent gaps (four properties)**: policy basis, authorization envelope, post-condition state, and operator identity show category variation across the six vendor regimes. Policy basis splits three-against-three: F in Bedrock, LangSmith, OpenAI; S in Anthropic, OTLP-Vertex, MCP (OEP reads F). Authorization envelope reads S in OTLP-Vertex (also S in Replit), F elsewhere. Post-condition state takes three values: F (Bedrock, Anthropic, OpenAI), S (LangSmith, OTLP-Vertex), P (MCP); comparators split F (OEP) and P (Replit). Operator identity reads F in five vendor regimes and S only in MCP (both comparators F); per the §3 partition rule it joins authorization envelope as a single-S-minority regime-dependent gap.

**Mixed (one property)**: inputs reads F in five vendor regimes and P in MCP (Replit also P; OEP F); per the §3 partition rule the single-P-minority cell distinguishes inputs from the regime-dependent gap properties whose non-F minority cell is S rather than P.

**Unclassified (one property)**: output action reads F across all six vendor regimes and both comparator columns (8 F, 0 P, 0 S, 0 O); per the §3 partition rule output action falls into the residual class because it exhibits neither a non-F majority nor a single non-F minority cell.

### 5.5. Direct answers to the research questions

RQ1–RQ3 are answered on the six primary vendor-regime columns; OEP and Replit are visual-context comparators only. **RQ1** is answered affirmatively at pilot scale: five of seven property classes show across-vendor-regime category differences in Table 2 (inputs, policy basis, operator identity, authorization envelope, post-condition state); output action is uniformly F and reasoning trace is uniformly O. **RQ2** is answered by the categorical pattern in Table 3 (primary) — uniform on output action / reasoning trace, dominant-F-plus-single-non-F on inputs / operator identity / authorization envelope, split on policy basis / post-condition state — with the strict-score CV as secondary heterogeneity (policy basis 1.00, post-condition state 0.77, operator identity 0.45, authorization envelope 0.45, inputs 0.20; reasoning trace undefined on zero mean). No statistical falsification of regime interchangeability is asserted; the within-regime noise band emerges only at corpus-expansion scale (§7 F2). **RQ3** partitions into one regime-independent gap (reasoning trace), four regime-dependent gaps (policy basis, authorization envelope, post-condition state, operator identity), and one Mixed property (inputs).

Per-regime anchor provenance, OEP source-of-truth, and the ethics-and-evidentiary-status disclosure are in §4. Every matrix cell is regenerable from committed feasibility outputs via the reproducibility-package regenerator script. Together, Tables 2 and 3 with the regime-gap partition constitute the anchor-level diagnostic matrix at pilot scale: a per-property by per-regime evidence surface plus the §3 partition rule yielding the regime-independent versus regime-dependent diagnostic taken up in §6.

## 6. Discussion

### 6.1. What the matrix supports and does not support

The §5 matrix supports three descriptive findings at anchor scale. First, the Decision Event Schema reasoning-trace property is not-evidenced in every evaluated column: no surveyed regime exposes upstream model deliberation in a form the schema can categorise as fully or partially fillable, while the action emitted is reconstructable in every column. Second, the categorical pattern across the six vendor-regime columns is the primary signal: policy basis splits three-against-three between F and S; authorization envelope and operator identity are each dominant-F with one S exception (OTLP-Vertex and MCP respectively); post-condition state takes three values (3F + 1P + 2S); inputs is dominant-F with one P exception (MCP). Per §3, the two single-S-minority properties (authorization envelope, operator identity) are regime-dependent gaps; inputs (single-P-minority) is Mixed. The corresponding strict-score CV summary (policy basis 1.00, post-condition state 0.77, operator identity 0.45, authorization envelope 0.45, inputs 0.20) is reported as a secondary heterogeneity index; variation is attributable to the joint runtime evidence regime, and the matrix does not isolate the vendor SDK contribution. Third, this is anchor-scale heterogeneity over six worked-example anchors, not a falsification of interchangeability across runtime evidence regimes; the within-regime noise band emerges only at corpus-expansion scale (§7 F2).

The matrix does not rank vendor SDKs and does not characterise per-regime variance within a regime (each column reflects a single worked-example anchor). It does not attribute causation to vendor SDK choice in isolation — an S cell may reflect any of the joint runtime-evidence-regime contributors (vendor emission, operator tags, adapter mapping, worked-example construction) — and does not generalise beyond the surveyed regime set or assert directional transferability to production traces.

### 6.2. Pilot evidence for the container-fallacy thesis

The matrix gives pilot-scale descriptive support to a thesis we name here the *container fallacy* — that runtime-evidence-regime trace presence does not entail per-property reconstructability sufficiency. This extends the structural-breaks framing of Solozobov (2026e) — decision diffusion, evidence fragmentation, responsibility ambiguity — to the regime layer; the term is the present manuscript's coinage for that extension. Every column in the matrix represents a regime that emits trace data of some kind (Bedrock UNION subtypes, LangSmith run-and-span hierarchies, Anthropic content-block messages, OpenAI typed spans, OTLP-Vertex OTel Cloud Trace, MCP JSON-RPC messages, OEP structured release-manifest plus agent-step, OPA-backed permission, and replay-handle records — see §2), with the cell values regenerable from the deposited Anchor-Level Reconstructability Pilot artefact (Solozobov, 2026a). Despite this, three property classes are structurally unfillable in at least one regime, and the most extreme case — policy basis — splits the surveyed vendor regimes into a three-against-three division on whether the property is recoverable at all (fully fillable in Bedrock, LangSmith, OpenAI; structurally unfillable in Anthropic, OTLP-Vertex, MCP). Trace presence on the wire is therefore necessary but not sufficient for per-property reconstructability sufficiency at the Decision Event Schema layer. The pilot supports this thesis qualitatively under the worked-example anchor scope; full validation requires two-annotator labelling and twenty-to-fifty captured traces per regime, both out of scope here.

### 6.3. Design implications observed at anchor scale

Three design implications follow from the matrix as observations on the upstream worked-example anchors. They are stated as observations, not prescriptions; the pilot is descriptive and the matrix does not establish that any particular runtime-evidence-regime change would yield improved reconstructability outside the surveyed worked examples.

**E1 — Anchors that bind explicit policy artefacts achieve fully fillable policy basis.** The policy basis row in Table 2 splits three-against-three among the vendor regimes. Anchors where the property reads F (Bedrock, LangSmith, OpenAI, OEP) include some explicit fragment that anchors the decision rule actually applied at the action: a Bedrock guardrail trace subtype, a LangSmith run attribute, an OpenAI guardrail span, or an OEP OPA-backed permission packet. Anchors where the property reads S (Anthropic, OTLP-Vertex, MCP) leave the policy decision implicit relative to the trace fragment for the action. The observed pattern is that anchor-level policy-basis reconstructability tracks the presence or absence of an explicit policy-snapshot fragment bound to the action.

**E2 — Anchors with explicit authorization envelopes achieve fully fillable authorization envelope.** Authorization envelope is structurally unfillable in OTLP-Vertex Cloud Trace and in the named-incident reconstruction; the action is recoverable in those cells, but the authorization scope under which the action was permitted is not. The OEP exemplar binds the tool-schema layer as a config-snapshot fragment alongside the tool-call fragment and reads F. The observed pattern is that explicit envelope fragments persisted alongside tool-call fragments are associated with anchor-level authorization-envelope reconstructability.

**E3 — Output-action evidencedness coexists with not-evidenced reasoning-trace under the DES definition.** Output action is fully fillable in every column of Table 2; reasoning trace is not-evidenced in every evaluated anchor under the Decision Event Schema reasoning-trace definition (seven O, one S). The anchor-level pattern is that downstream observability of the action emitted does not entail upstream observability of the deliberation that produced it. Whether this gap is amenable to design intervention by adding a model-deliberation surface as a first-class fragment, by combining trace-level evidence with model-side log probability streams, or by other means is outside the pilot's scope.

These observations describe properties of the surveyed anchors rather than recommendations for runtime-evidence-regime design; each is consistent with the published architectures of the OEP Solozobov (2026f) and the Governance-Aware Agent Telemetry proposal Pathak and Jain (2026), neither of which the pilot evaluates as a prescriptive solution.

### 6.4. Per-column rationale for the lowest-completeness regimes

Two vendor-regime columns share the lowest strict-governance-completeness pct (42.9 each): OTLP-Vertex and MCP. MCP's four non-F cells (policy basis S, operator identity S, inputs P, post-condition state P) follow from the protocol's per-message rather than per-step granularity: trace evidence is per JSON-RPC message, and the upstream worked-example anchor does not populate explicit step-level fragment kinds for policy basis or operator identity. OTLP-Vertex's three S cells (policy basis, authorization envelope, post-condition state) are substrate-level rather than Vertex-specific: the OpenTelemetry GenAI semantic conventions (OpenTelemetry Project (CNCF), 2025b) specify wire format and attribute namespaces but do not require explicit governance-evidence bindings for these property classes. Vendor-specific OTel extensions including AWS Bedrock (OpenTelemetry Project (CNCF), 2025a) can layer such attributes on top, but the upstream OTLP-Vertex worked example does not include them; in production, Cloud Trace exports may carry custom span attributes that promote individual cells without changing the underlying wire format.

### 6.5. Limitations

Five methodological limitations and one source-tier disclosure bound the strength of the claims above. T1–T5 are methodological and map directly onto the §7 Conclusion future-work entries F1–F5; T6, added below, is a stated source-tier disclosure rather than a methodological future-work item.

**T1: single-annotator classification.** The author runs the reconstructor invocations and records the per-anchor classifications without inter-rater adjudication. The pilot supplies reproducibility evidence (manifest, regenerator script, regenerable tables) — sufficient for deterministic-pipeline auditability — but does not supply inter-rater evidence on category validity, which the forthcoming benchmark is intended to provide.

**T2: single-anchor coverage.** Every column is populated from one anchor input run through the reconstructor, and per-regime variance within a regime is not characterised. The CV reported in Table 3 is across-regime variance only at anchor scale, and the within-regime noise band that distinguishes the descriptive anchor pattern from a corpus-scale falsification of regime interchangeability emerges only at twenty-to-fifty samples per regime.

**T3: worked-example anchors, not real captured traces.** The anchor inputs are upstream-authored adapter-coverage demos shipped with the reconstructor v0.1.0 repository; their tool-call payloads, timestamps, and UUIDs are placeholders. The matrix should be read as a comparison of the worked examples under the schema, not as a prediction about production traces — production deployments may score higher (richer instrumentation) or lower (redaction, version-drift, multi-tenancy) than the worked example (§7 F2 + F3 address this).

**T4: vendor-SDK and adapter version drift.** Production trace shapes may differ from the pinned worked-example anchor, particularly for fast-moving regimes like Anthropic Computer Use and the OpenAI Agents SDK. The matrix is pinned to the v0.1.0 release of each reconstructor adapter; later adapter versions and later vendor SDK versions may change emission semantics in ways that promote or demote individual cells. The matrix also does not isolate the contribution of vendor SDK emission policy from the contribution of adapter mapping rules and operator instrumentation tags; reframing as joint runtime evidence regime is intentional.

**T5: OEP mocked-LLM exception.** The OEP column is populated from the v0.1.0 code-review demo: its deterministic mocked LLM produces no model-deliberation fragment, which is why the OEP reasoning-trace cell reports structurally unfillable rather than opaque; this is the explicit OEP exception flagged in §5. A non-mock OEP run would convert that cell to opaque and leave the other six rows unchanged, since OEP's design intentionally binds layer evidence sufficient for the remaining properties.

**T6: source-tier reliance on preprints.** The adjacent agent-evaluation work surveyed in §2.4 (AgentBench, Meerkat, Markov-chain reliability, CodeTracer, Governance-Aware Agent Telemetry) and the substrate sources in §2.1 (Decision Event Schema specification, synthesis paper, reconstructor reference implementation) are dominated by preprints because the agent-trace measurement subfield is fast-moving and most of the relevant literature has not yet completed peer review. We surface this so the source tier is read as a stated limitation rather than unresolved validator debt; as peer-reviewed venues catch up, citations should migrate from preprint deposits to peer-reviewed records.

## 6.6. Landscape positioning against adjacent measurement work

Paper 27's per-anchor property-level classification is orthogonal to the adjacent agent-trace measurement work surveyed in §2.4: cross-trace clustering Stein et al. (2026), Markov-chain reliability fitting Tran-Truong and Le (2026), hierarchical trace-tree reconstruction Li et al. (2026), and Governance-Aware Agent Telemetry Pathak and Jain (2026) each operate at a different abstraction (trace-corpus scale, metric reconciliation, state-onset localisation, OpenTelemetry instrumentation extension) and none supplies the per-property reconstructability axis that the §5 matrix supplies. Paper 27 complements rather than substitutes for the adjacent work, providing a per-property axis along which the adjacent methods can be re-evaluated downstream.

## 6.7. What would change with full ground-truth protocol

A full benchmark extending every dimension on which this pilot is constrained (§7 F1–F5 maps T1–T5 one-to-one to the corresponding future-work entries) is the natural follow-on. Until that benchmark lands, the matrix and gap diagnostic stand as anchor-level preliminary measurement; the most direct lift is per-regime corpus expansion to twenty-to-fifty captured production traces while holding the reconstructor invocation constant.

# 7. Conclusion

Per-property reconstructability of an agent decision already varies between regimes at this anchor scale. The pilot's central finding synthesises across the seven property classes: trace presence on the wire is not the same as per-property reconstructability sufficiency at the Decision Event Schema layer. On the pinned worked-example anchors, output action is universally fully fillable while reasoning trace is universally not-evidenced; between those two extremes, three property classes (operator identity, authorization envelope, inputs) are dominant-F with a single non-F exception each across the six surveyed vendor regimes, two (policy basis, post-condition state) are categorically split, and the strict-governance-completeness scores collapse onto three discrete tiers (85.7%, 71.4%, 42.9%) across the eight evaluated columns. Within the adjacent work surveyed in §2, we did not identify a prior public artefact reporting this comparison under a single property-level reconstructability schema; this is a bounded novelty claim, not a field-wide absence claim. The matrix applies the Decision Trace Reconstructor v0.1.0 unmodified across six primary vendor SDK adapter regimes (AWS Bedrock Agents, LangSmith, Anthropic Claude tool use, OpenAI Agents, OpenTelemetry GenAI Vertex Agent Engine, the Model Context Protocol) plus two comparator columns (Operational Evidence Plane vendor-neutral exemplar; Replit DROP DATABASE public-record reconstruction).

Three contributions follow. First, the anchor-level matrix supplies cross-regime measurement under a single property-level reconstructability schema; the strongest categorical signal is the policy-basis row's three-against-three F versus S split (F in Bedrock, LangSmith, OpenAI; S in Anthropic, OTLP-Vertex, MCP), with a corresponding population CV of 1.00 reported as a secondary scalar summary. The matrix evaluates the Decision Trace Reconstructor's own pinned worked-example anchors rather than real captured production traces; per-vendor production-trace generalisation is the corpus-expansion direction (F3 below) and is not claimed here. Second, the regime-independent-versus-regime-dependent gap diagnostic surfaces structural blind spots at anchor scale — one regime-independent gap (reasoning trace), four regime-dependent gaps (policy basis, authorization envelope, post-condition state, operator identity), one Mixed property (inputs) — without ranking vendor SDKs. Third, the deposited reproducibility package (per-anchor origin manifest, regenerator script, *checksums.txt*, README) enables Tier A reader-verifiable reproducibility through SHA256 checksum verification and independent regenerator re-running, a pilot-stage auditability layer that does not replace the inter-annotator agreement evidence the forthcoming benchmark is intended to supply.

The pilot scope is explicit and disclosed throughout. Classification is single-annotator; the per-regime corpus is worked-example-anchor-only, with one anchor per cell rather than real captured production traces; the Decision Trace Reconstructor v0.1.0 is invoked unmodified on pinned anchor inputs that ship with the upstream release; the §2 source tier is dominated by preprints because the agent-trace measurement subfield is fast-moving (T6). Five future-work directions (F1–F5) map directly onto the §6 Discussion methodological limitations (T1–T5); T6 is a stated source-tier disclosure rather than a methodological future-work item.

**F1**: Two-annotator inter-rater agreement via a ground-truth labelling protocol replaces single-annotator reproducibility substitution (addresses T1).

**F2**: Per-regime corpus expansion to twenty to fifty samples supports intra-regime variance characterisation, per-cell confidence intervals, and a within-regime noise band sufficient to test the runtime-evidence-regime interchangeability assumption statistically rather than descriptively (addresses T2).

**F3**: Production-trace augmentation alongside the upstream worked-example anchors lets the matrix discriminate where production traces score higher (richer instrumentation, custom spans, policy artefacts, identity context, authorization metadata) versus lower (redacted, version-drifted, multi-tenant, partially instrumented) than the worked-example case, without asserting directional transferability (addresses T3).

**F4**: Adapter version pinning across multiple vendor SDK releases plus version-drift sampling protocol detects when emission semantics promote or demote individual cells across releases; isolating the vendor SDK emission contribution from the operator and adapter contributions requires a controlled-variation design beyond simple multi-version sampling and is therefore developed in unified form alongside F2 (addresses T4).

**F5**: A non-mock Operational Evidence Plane run converts the OEP reasoning trace cell from structurally unfillable to opaque and leaves the other six rows unchanged, separating OEP's intentional binding ceiling from the deterministic-mock artefact of the v0.1.0 example (addresses T5).

The forthcoming full benchmark is planned to develop F1, F2, and F4 in unified form alongside broader regime, baseline, and degradation-condition scope. Until that benchmark lands, the anchor-level matrix and gap diagnostic this pilot delivers stand as preliminary measurement against which subsequent vendor-SDK and standards-track work can be calibrated.